# Design and expression of TolC as a recombinant protein vaccine against *Shigella flexneri* and evaluation of immunogenic response in mice

**Razieh Veisi**

Shiraz University of Medical Sciences, Veisir@sums.ac.ir

**Nahal Hadi**

Shiraz University of Medical Sciences, Hadina@sums.ac.ir

**Shahram Nazarian**

Shahed University

**Javad Fathi**

Shiraz University of Medical Sciences

## Abstract

**Background:** *Shigella* is one of the main causes of dysenteric diarrhea, which is known shigelosis. Shigelosis causes 160,000 deaths annually of diarrheal disease in a global scale especially children less than 5 years old. No licensed vaccine is available against shigelosis, therefore, efforts for development the effective and safe vaccine against *shigella* as before needed. The reverse vaccinology (RV) is a novel strategy that evaluate genome or proteome of the organism to find a new promising vaccine candidate. In this study, immunogenicity of a designed-recombinant protein is evaluated through the insilico studies and animal experiments. **Methods:** In the first step, proteome of *shigella flexneri* was obtained from UniProtKB and then the outer membrane and extracellular proteins were evaluated. Ultimately, TolC as an outer membrane protein was selected. In next steps, confirmation of pre-selected protein was performed in order to transmembrane domains, homology, conservation, antigenicity, solubility, and B- and T-cell prediction by different online servers. **Results:** TolC as a conserved outer membrane protein, using different immunoinformatics tools had depicted acceptable scores and was selected as an immunogenic antigen for animal experiments studies. Recombinant TolC protein after expression and purification, was administered to BALB/c mice intraperitoneally over three routes. The sera of mice was used to evaluate the IgG1 production assay by indirect-ELISA. The immunized mice displayed effective protection against 2 × LD50 of *shigella flexneri* ATCC12022 (challenge study). **Conclusion:** Therefore, the reverse vaccinology approach and experimental results demonstrated that TolC as a novel effective and immunogenic antigen is capable for protection against shigellosis.

**Keywords:** In silico; Reverse vaccinology; *Shigella*; Diarrhea; Recombinant vaccine

**Introduction**

According to latest Global Burden of Disease estimates *Shigella* is one of the five main pathogens causing diarrheal disease and is major cause of deaths by the year through the diarrhea. *Shigella* as an enteric pathogen is one of the main causes of moderate to severe diarrhea and contributes to significant morbidity and mortality, particularly in younger children in developing countries (1). Transformed cell lines and animal models have been widely used to investigate the pathogenesis of *Shigella*. Shigellosis initially involves the epithelial layers via M cells and enters into cells by the type three secretory systems (TTSS) and leads to the destruction of the epithelial layers and causes abscesses, ulceration, watery diarrhea, and abdominal cramps symptoms. In the absence of effective treatments, shigellosis may develop severe secondary complications (2, 3).

Following these symptoms, bacterial invasion into the epithelial layers induces acute inflammation and subsequent tissue destruction. Shigellosis is caused by four species of *Shigella* including *S. flexneri* is the most frequently isolated species in developing countries; *S. sonnei* is more common in low- and middle-income countries; and *S. dysenteriae* has historically caused epidemics of dysentery, *S. boydii*, a cause of infection in less-developed countries (2, 4).

In addition, increasing antibiotic resistance means that the development of a vaccine preventing *Shigella* infections remains a high priority for the World Health Organization (5). Over the past decades, the production of ESBL and AmpC beta-lactamases has created problems for the antibiotic treatment of shigellosis. Consequently, the production of various forms of these enzymes can be considered the main mechanism of resistance to beta-lactam agents, i.e., the cephalosporins (6, 7).

Various vaccination strategies have been used to prevent shigellosis, However, the majority of vaccine candidates assessed against *Shigella* have a low immune response, and Thus far, no licensed prophylactic. Or therapeutic vaccine against shigellosis is commercially available (8).

In the recent decade, the enhancement of numerous biological databases and computational tools, in silico algorithms strategies, and statistical methods, immunoinformatics tools expanded the sequence-based approaches designing in vaccine research and development. The use of conversed protein antigens among the various serotypes of *Shigella* strains is evaluated as a state-of-theart strategy (9, 10).

Immunoinformatics has impelled the rising of vaccinomics by providing computational biology servers as a new way to generate an applicable vaccine candidate (11). Recently, the concepts of the immunoinformatics approach have combination with reverse vaccinology (RV) to design an effective vaccine for the investigation of the immune system. Reverse vaccinology was first described by Rino Rappuoli in 2000 as using genetic data as a starting line in vaccine design, instead of the reliance on the pastime of the pathogen itself (12).

Reverse vaccinology was used successfully against Neisseria meningitidis serogroup B (MenB) as a universal vaccine (13). In this study, TolC protein as an efflux pump protein is evaluated to determine immunogenicity against *S. flexneri* in the mouse. Using the immunoinformatic tools an immunogenic region of this protein was designed and produced by the expression vector, Pet28a (K+). After that used as a vaccine in mouse models. TolC and the other membrane protein contain outer membrane factor (OMF), these form outer membrane proteins forming trimeric channels in Gram-negative bacteria. TolC as an outer membrane protein is the tip structure of the AcrAB-TolC pump in gram-negative bacteria and this protein involves in increasing multidrug resistance and antibiotic resistance mechanisms such as quorum sensing, biofilm formation, two-component regulons, etc. in bacteria. TolC has an important role in increasing multidrug resistance and secretion of toxins involved in bacterial infection (14). The in silico studies of TolC recombinant protein were considered by immunoinformatics and proteomic approaches. In the current study, the rTolC protein of *Shigella spp* was considered an immunogenic antigen for the evaluation of immunogenicity and protectivity against mice.

## Materials and methods

### In silico studies

An in silico study was performed to design a recombinant protein through the reverse vaccinology (RV) approach.
A typical RV pipeline occurs in silico and involves screening the whole protein-coding sequences of the genome of a pathogen (15). In the first, the proteome of *S. flexneri* Serotype (ATCC 12022) has been retrieved from UniProtKB in FASTA format after which imported to PSORTb v.3.0.2 online server for subcellular localization. PSORTb v.3.0.2 is a specific and wide-used online server that is applied for the identification of recognized organic functions associated with proteins' subcellular localization (16).

To determine the annotation level for UniProt genes, the outer membrane proteins as potential candidates have been selected with an 'annotation score' ≥ 9 for the next analyses.

Surface-uncovered proteins, because of their near contact with host cells, and therefore immune system stimulations are appropriate subunit vaccine targets. Thus, on this examination and as a starting step, TolC as an outer membrane protein was selected by this server to next evaluations.

Cloning and expression of proteins are needed for the evaluation of transmembrane domains. In our analyses, the protein with transmembrane domains < 1 was potent. CCTOP v. s.1.1.0 was used for searching hydrophobic amino acids and a maximum divergence of amino acid composition, respectively. Default values of their parameters were used to determine the number of transmembrane domains in TolC as an outer membrane protein. TolC protein was blasted on the BLASTp server (NCBI database) to investigate the auto-immune response of proteins through the determination of homology with the mouse (taxid: 10088) and human (taxid: 9606) proteins (17, 18). Also, this server was accomplished against non-redundant protein sequences (proteins with identify and query cover < 30% were approved as non-homologous proteins). A vaccine candidate should be a wide-spectrum vaccine and be used in various species and strains. Also, protein conservation should be determined (The query cover more than 80% was as cut-off value) (19).

The antigenicity of the TolC protein was computed using the Vaxijen v. 2.0 server (The Edward Jenner Institute for Vaccine Research, Compton, UK). Also, ANTIGENpro for more reliable results was used (20).

*E.coli* as the expression host might result in insoluble and soluble proteins, the solubility evaluation is important because the recombinant protein must be soluble to avoid inclusion body forming. Here, the TolC sequence was analyzed with protein-sol server to observe the prospective solubility (20). In silico prediction of B-T cell epitopes were performed for the TolC of *S. flexneri* (GenBank accession no. CCO02501.1) through online servers. To increase prediction reliability, different servers with sundry methods were used.

The Immune Epitope Database (IEDB) for both B-T cells, abcPred by recurrent neural network method was used to calculate the density of epitope based on protein length for B cell, and MHCPred v.2.0, based on a quantitative structure-activity relationship (QSAR) method for T cell are another epitope prediction tools (21). MHCPred was used for the prediction of strong binder of T-cell epitopes (IC50 < 50 nm). Which were provided by the server. The results of these steps and the other parameter were listed, and after the final approval, the experimental phase was performed. Applied servers in this study were exhibited in Table 1.

Table 1 Online servers.

| Function | Server | Website |
| --- | --- | --- |
| Reference proteome | UniProtKB | https://www.uniprot.org/ |
| Subcellular localization | PSORTb v. 3.0.3 | https://www.psort.org/psortb/ |
| Transmembrane domains | CCTOP v. s.1.1.0 | http://cctop.ttk.hu/ |
| Homology analyses | BLASTp | https://blast.ncbi.nlm.nih.gov/Blast.cgi?PAGE=Proteins |
| Antigenicity | VaxiJen v2.0 | http://www.ddg-pharmfac.net/vaxijen/VaxiJen/VaxiJen.html |
| Antigenicity | ANTIGENpro | https://scratch.proteomics.ics.uci.edu/ |
| Solubility | Protein-Sol | https://protein-sol.manchester.ac.uk/ |
| Continuous B-cell epitope | ABCpred | http://www.webs.iiitd.edu.in/raghava/abcpred |
| Linear B-cell epitope | IEDB B-cells | http://www.tools.iedb.org/main/tcell |
| T-cell epitope (MHC class II) | IEDB T-cells | http://www.tools.iedb.org/main/tcell |
| T-cell epitope (MHC class I) | MHCPred v.2.0 | http://www.ddg-pharmfac.net/mhcpred |
| Codon optimization | GenScript | https://www.genscript.com/gensmart-free-gene-codon-optimization.html |

**Mice**

Six-week-old female BALB/c mice (25-30 gr) were provided by the Center of Comparative and Experimental Medicine Shiraz University of Medical Sciences, Shiraz, Iran. In this study, animals were maintained in this center, and keeping them was provided with the procedures described in the guide for the care and use of laboratory animals. They were handled in all experiments while using standard conditions approved by the Bioethics and Safety Committee of the Faculty of Bacteriology and Virology department. Food and water conditions were the same for all groups and Specific pathogen-free (SPF).

**Bacterial strains and plasmid**

*S. flexneri* Serotype ATCC 12022 (generous gift of Dr. B. Pourabbas T, Alborzi Clinical Microbiology Research Center, Shiraz, Iran) and *E.coli DH5α* (Iranian Research Organization for Science and Technology (IROST), Tehran, Iran) were used in this study. In the transformation step, *E.coli DH5α* as a competent cell was mixed with plasmid construction and a total of 250 μl fresh LB medium was added for cell recovery at 37°C for 2 h. The recovered cultures were spread into the LB plate with the indicated antibiotic (70 μg/mL kanamycin) (22).

**Construction of tolC**

The *tolC* gene, devoid of signal peptide and codon-optimized for expression in *E. coli* was synthesized by GenScript (https://www.genscript.com/gensmart-free-gene-codon-optimization.html). The gene was subcloned into the pET28a(+) vector, between the EcoR1 and XhoI sites. Recombinant TolC spanned 494 amino acids. This interest sequence was cloned in the designed construction vector (23). The plasmid construction was transformed into competent *E. coli DH5α* cells, and transformants were selected on LB agar plates containing kanamycin (70 μg/ml).

**Expression and purification of recombinant TolC protein**

The *tolC* sequence containing His-tagged residues was expressed in *E. coli BL21* (DE3). Briefly, an overnight culture of the *E. coli BL21* (DE3) cell in LB medium with kanamycin was allowed to grow until the culture density reached 0.6 (OD600). The expression was induced using 1 mM isopropyl β-D-1-thiogalactopyranoside (IPTG), and allowed to grow for 4 h post-induction. The suspension from the former step was centrifuged at 6000 g at 4ºC for 10 min. cells were lysed by buffer containing Tris, Imidazole, NaCl, pH 7.5, and Kept under 4ºC for 10 minutes. The centrifugation of suspension was performed at 6000 g at 4 C for 8 min. Protein products were separated using SDS-PAGE through size and purification determination (23, 24).

After expression, the protein was purified by affinity chromatography using a Ni-NTA Agarose matrix (Sigma Aldrich). Briefly, to remove insoluble recombinant proteins from suspension, after sonication and centrifugation, resuspended in 1X binding buffer (0.5 M NaCl, 20 mM Tris-HCl, 5 mM imidazole, pH 7.9) containing 6 M urea. After 45 minutes of incubation on the ice, the suspension was centrifuged, and the supernatant passed through a 0.45 μm filter and dialysis against PBS.

Followed by purification, proteins were evaluated by SDS-PAGE and coomassie blue staining. The expressed recombinant protein was confirmed through western blot by probing with IgG primary anti-His×6 and secondary anti-IgG HRP conjugated antibodies (Sinaclon, Iran). Ultimately, The concentration of total protein was quantified by the BCA Kit ( Bicinchoninic Acid Assay, Thermofisher ) (23).

**Mice immunization**

Six-week-old female BALB/c mice ( Center of Comparative and Experimental Medicine Shiraz University of Medical Sciences, Shiraz, Iran) were allocated into three groups including group 1 with 12, group 2 with 12 and group 3 with 12 mouse for the immunization step, 5 groups (4 mouse/ group) for challenge step, and 5 groups (6 mouse/ group) for LD50 determination. In the injection step, mice were immunized subcutaneously with 125 μl of rTolC (25μg/mouse, group 1) with 125 μl adjuvant (Sigma, USA), 125 μl rTolC (25 μg/mouse, group 2) and 125 μl phosphate buffer saline with 125 μl adjuvant (group3). The injection steps (first and second booster) were performed based on below graphical process (Fig 1).

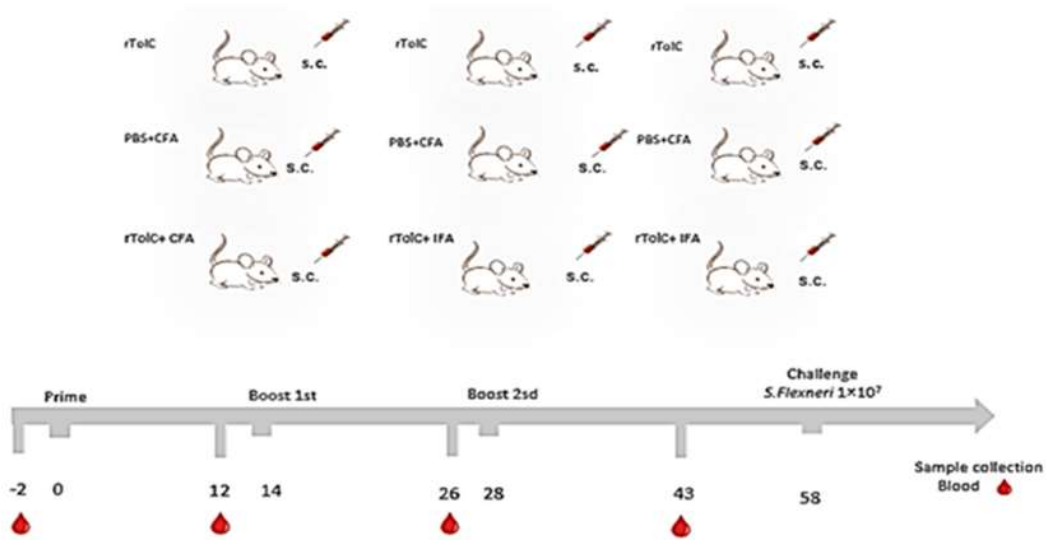

**Fig 1** Immunization and sample collection process. Days of vaccine routes injections, blood drawing, adjuvants and challenge step are illustrate. Beriefly, Administration was applied over main injection and two boosters. Blood collection is showed in red droplet. CFA, compelet Feround's adjuvant. IFA, incomplete Feround's adjuvant. S.b., subcutaneous.

**LD50 determination**

The lethal dose of *S. flexneri* Serotype 2b (ATCC 12022) was evaluated to determine the protective efficacy of the candidate recombinant protein in immunized mice through serial dilution, plating and counting. The challenge step was performed 30 days after the last immunization. For more details, after growth of *S. flexneri* in TSA at 37 °C and then colonies were grown in TSB at 37 °C overnight to reach OD650 = 1 (based on the standard curve of bacterial growth in this absorption exist $10^9$ CFU/ml). Serial dilution through the standard curve of bacterial growth was provided from colonies (low= $10^5$ CFU/ml to high= $10^9$ CFU/ml) to LD50 determination. 5 groups of mice were injected by different lethal doses of *S. flexneri*. Finally, the lethal dose of *S. flexneri* was determined $1\times10^7$ CFU/mouse (22).

**Determination of Specific Antibodies in Peripheral Blood**

Two days before each immunization and 15 days after the third immunization, Blood was drawn through the tail and collected to evaluate the humoral immune response. After the centrifugation, sera were isolated for specific IgG detection by Indirect Elisa. Briefly, 96-well plates were coated with 1μg of the rTolC protein diluted in 200 μl of coating buffer (pH 9.6). after incubation at 4 °C overnight, washing was performed three times with PBST (PBS-0.05% Tween-20) and then wells were blocked with blocking buffer (PBS 0.8% gelatin) for 2 h at 37 °C. The wells were washed with PBST three times and then were incubated with diluted serum at 37 °C for 2 h. following this step, wells were washed three times with PBST, and plates were incubated at room temperature with 100 μl secondary antibody goat anti-mice IgG (1:2000) conjugated to horseradish peroxidase (Serotec, Oxford, UK). 1 h. After adding 200 μl of the substrate TMB (3, 3′, 5, 5′-Tetramethylbenzidine)/ H2O2 (Thermo Fisher Scientific, Waltham, MA, USA) was incubated in the dark for 1 h at room temperature. In the last step, 50 μl of 2 M H2SO4 as a stop solution and was read the optical density at 450 nm by plate reader. At the room temperature (RT), the reaction was stopped by adding 50 μL H2SO4 2N. The plates were read at OD450 nm using VictorX3 (PerkinElmer, Waltham, MA, USA). The antibody titers were expressed as mean standard deviation (SD) of log10 of the last reciprocal serum dilution above the cut-off. The cut-off values were calculated according study that performed by Frey et al (25).

**Challenge studies**

The lethal dose of *S. flexneri* serotype 2b (ATCC 12022) was measured at $1\times107$ CFU/mouse. For the challenge and survival step, Two bacterial doses including $2 \times$ LD50 and $4 \times$ LD50 were administered to the test and control groups of vaccinated mice (n = 4 mice/group). 25 μl of bacterial suspension was injected through intraperitoneal (IP). The results followed up to evaluate the mortality rate among mice one 15 days after challenge step.

**Statistical analysis**

The Graphpadprism software (version 9.4.1) was used to generate all graphs and statistical analyses. Specific antibodies were analyzed by the differences between antibody response of immunized and non-immunized mice were analyzed by One-way repeated analysis of variance (ANOVA). The protection studies were analyzed using Kaplan-Meier survival curves. And p-value of 0.0001 or less was considered to be significant for all tests.

**Results**

**Subcellular localization and topology analyses**

The proteome of *S. flexneri* (serotype *2b*, strain *12022*, UP000029282) which is a highly virulent strain, was obtained from the UniProt database. The full proteome of *S. flexneri* contained 4,725 proteins. Extracellular proteins and outer membrane proteins of the *S. flexneri* proteome were analyzed by the PSORTb server. TolC protein as an outer membrane protein with a full score of 10 was predicted for the next steps. Through the CCTOP server, the number of the transmembrane domain of TolC was zero. (Table 2)

**Conservation and homology analyses**

By the BLASTp servere analyses, TolC was a conserved protein among *Shigella* species and could be a cross-reactive protein agaist different strains. Also,TolC was confirmed a non-homologous protein with human and mouse proteome and it is the reason that the TolC protein will induce immunogenicity without auto-immune response and cross-reaction in host (Table 2).

**Antigenicity and solubility**

Vaxijen and ANTIGENpro were illustrated that the TolC protein has admissible antigenicity. By the Vaxijen and ANTIGENpro the antigenicity score of protein was higher than 0.5 and 0.9, respectively. Also, through the protein-sol program, the solubility of protein was predicted more than 0.45. Finally, According to the former results, TolC as an antigen outer membrane protein was selected against *S. flexneri* (Table 2).

Table 2  The step by step process to confirm TolC as a new effective immunogenic protein candidate.

| Feature | Program | Score |
| --- | --- | --- |
| Outer membrane predicted score | PSORTb | 10.00 |
| Transmembrane domain | CCTOP v. s.1.1.0 | 0 |
| non-conserved proteins | BLASTp | < 80% |
| host -homologous proteins | BLASTp | >30% |
| Antigenicity | Vaxijen | 0.54 |
| Antigenicity | ANTIGENpro | 0.92 |
| Solubility | Protein-sol | 0.45 |

**B- T cells epitope prediction**

According to results of Vaxijen and Antigenpro servers the sequence of TolC protein confirmed as an antigenic adequate candidate. The detailed results analyses using different servers. For B cell epitope prediction, Linear epitopes were predicted by IEDB B-epitopes, discontinuous B cell epitopes were predicted by abcPred based on artificial neural network (26), MHCPred v.2.0 was used to predict the binding affinity of major histocompatibility complex (MHC I) by the additive method (27), and IEDB T-epitopes for MHC class I.

*tolC* **sequence optimization**

Codon optimization was evaluated to enhance the translational function of *E. coli*. The encoding gene sequence of TolC protein was confirmed as an optimized sequence using GeneScript tool. The sequence with Codon Adaptation Index (CIA) 0.94, was in the acceptable range (Ideal value: 0.8-1.0), and also the GC content was 46.70%, that was located in range (Ideal value: 30%-70%).

**Gene Construction, expression, and purification of the recombinant TolC Protein**

The tolC gene was optimized (gene accession number: NP_838556), synthesized, and subcloned into pET28a (+) vector (Shingen, china). Expression of protein was induced with 1mM IPTG in *E.coli BL21-DE3*. The expression results were analyzed by 10% SDS-PAGE and western blot (Fig 2). The result of expression showed that this recombinant protein has a molecular weight close to 36.4 KD. The protein was purified under denatured conditions by Ni-NTA chromatography and then was confirmed by SDS-PAGE (Fig 3).

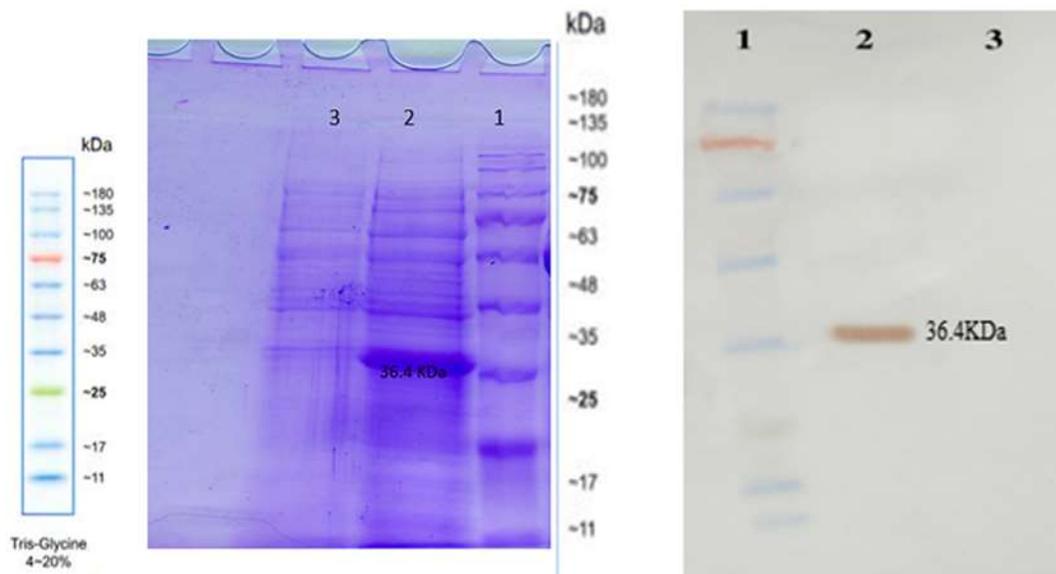

**Fig 2** [Left]: SDS-PAGE of expressed rTolC with 36.4 KD in E.coli BL21 cells after incubation with 1 mM IPTG for 4 h at 37 °C. Lan1: Protein marker (116 KD). Lane 2: Protein after incubation with IPTG. Lane 3: Protein before incubation with IPTG. [Right]: Western blot test confirmaed the presence of TolC protein after the purification-elution by Ni-NTA affinity chromatography. Lane1: Protein marker (180 Kda). Lane 2: rTolC (36.4 Kda). Lane3: Control (Bovine serum albumin).

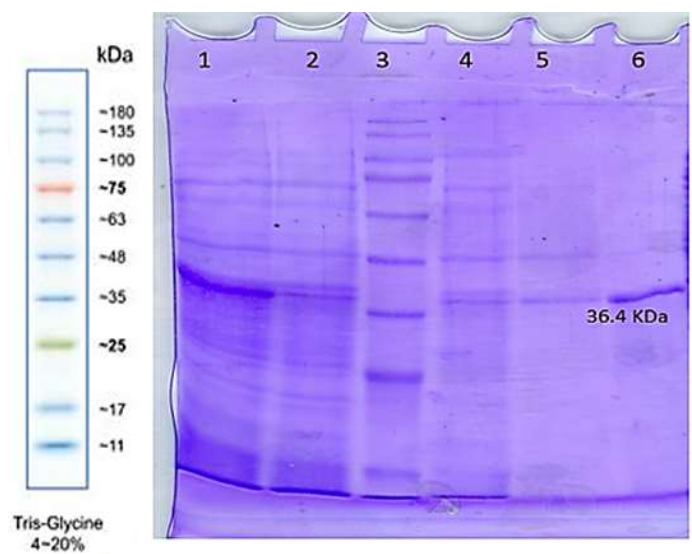

**Fig 3** SDS-PAGE after purification through affinity chromatography using Ni-NTA agarose matrix. Lane 1: Expression of rTolC after incubation with IPTG. Lane 2: Protein marker (116 KDa). Lane 3: Expression of rTolC before incubation with IPTG. Lane 4: Flow-through. Lane 5: Wash C fraction (pH=4.3). Lane 6: Wash D fraction. Lane 7: Elution fraction

**Immunogenicity evaluation in mice**

The specific antibodies of immunized mice were measured by ELISA test. Ten days after each immunization, sera were collected and antibody levels were determined. In the groups that immunized with rTolC and rTolC + adjuvant the level of antibodies after the first booster depict significantly increased in comparison with the control group. After the second booster the level of antibodies reached the highest levels in groups that immunized with the TolC antigen. Also, the antibody production in the control group was without a significant increase in the three immunization steps (Fig 4).

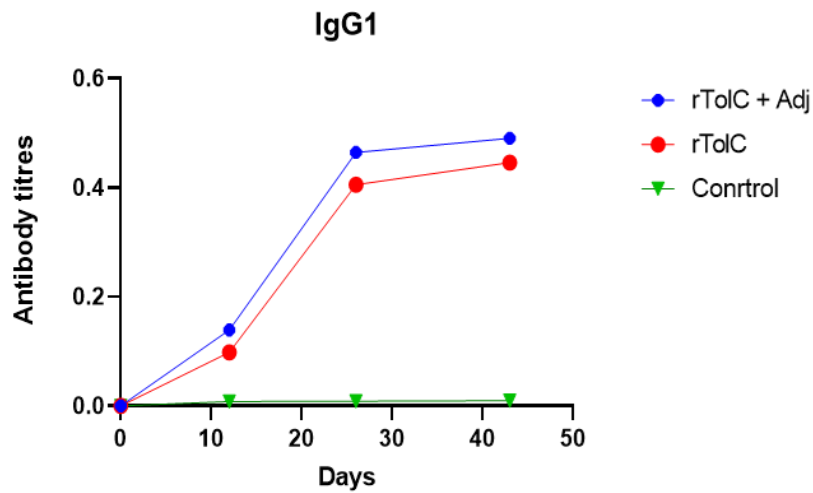

**Fig 4** Antibody serum titers in controls and immunized mice (IgG1). Control, rTolC, and rTolC + adjuvant groups were immunized through intrapritoneal route (i.p.). The first and two booster injections were applied in 1st, 14th, and 28th days respectively. Antibody titres were evaluation by one-way ANOVA test. The results (letters and symboles in above graph) of statistical analysis depicte significant values of $p < 0.0001$ compared with the both control and test groups.

**Protective response against *S. flexneri***

The high level of humoral responses was applied for a protective response against 2×107 LD50 of *S. flexneri*. In the group that was challenged with 2×107 LD50 Kaplan–Meier survival curves showed that mice group Subcutaneously immunized with rTolC + Adjuvant (Group A) depicts the highest levels of protection, with 90 % survival rate after 30 days ( IP injection of *S. flexneri*). Moreover, group B (immunized with rTolC) showed 70% survival rate after the challenge assay. However, the control group mice (PBS+Adjuvant) died within 4 days after infection (Fig 5). The immunized groups that were challenged with 4×107 LD50 of *S. flexneri* illustrted the lowest survival rate aproximatly 75% and 55% in group A and group B, respectively (Fig 5).

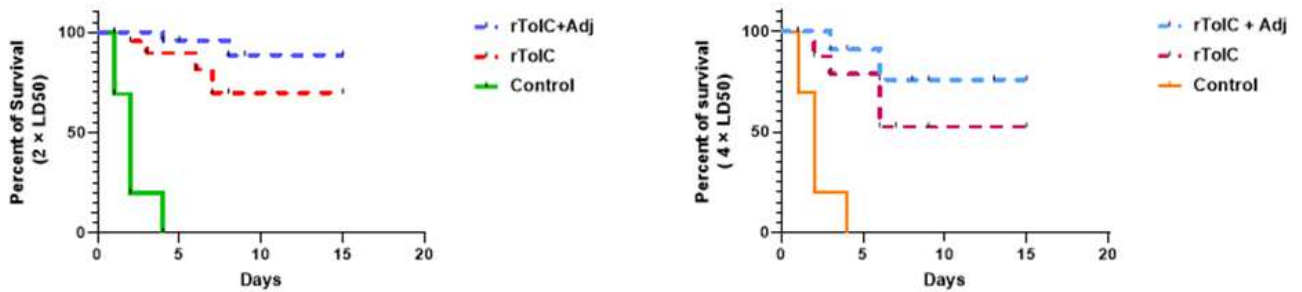

**Fig 5** Protective efficacy of recombinant TolC protein and evaluation of survival rate of immunized mice. Female BALB/c mice groups (n = 4 mice/group) were immunized through interapritoneal routes by 25 µg/mouse of rTolC and rTolC + adjuvant (Freund's Complete Adjuvant and Freund's incomplete Adjuvant). After last booster, control and test groups of mice were challenged with 2 doses of *S.flexneri* (2×107 CFU/mouse, 4×107 CFU/mouse) through i.p and countinuously mortaloty rate was followed up for 20 days. The groups of mice that were challenged with 2×10$^7$ CFU/mouse, the immunized mice with rTolC + adjuvant (group A) and rTolC showed 90% and 70% of survival rate and control group that recieved PBS + adjuvant died over 4 days after challenge. The groups of mice that were challenged with 4×10$^7$ CFU/mouse, the immunized mice with rTolC + adjuvant (group A) and rTolC showed 75% and 50% of survival rate and also, control group that recieved PBS + adjuvant died over 4 days after challenge. Survival rates were analyzed using the long-rank test ($P < 0.0001$).

**Discussion**

Shigellosis as a major cause of mortality in young children, elderly people and also increased multi-drug resistance is a global problem over the world (28). *S. flexneri* is one of the most important agents of shigellosis strain prompting the development of a safe and effective vaccine. In the current study, we used the reverse vaccinology method to verify and characterized an outer membrane protein of *S. flexneri* by evaluation of subcellular localization, transmembrane domains, homology with human and mice, conservation, antigenicity, solubility, and B-T cells epitope prediction features to develop a novel vaccine candidate.

In previous studies, using outer membrane proteins have been predicted and identified as the individual recombinant protein vaccine against gram negative bacteria (29-31) and *Shigella* species through bioinformatic tools (12, 17, 32) To design a vaccine candidate by reverse vaccinology approach, the first step is subcellular localization and topology analysis of organism proteome. Outer membrane proteins as surface proteins because of their direct interaction with the host cell are likely immunogenicity induction contrary to the inner membrane and cytoplasmic proteins (33).

These advantages and findings from novel individual protein vaccines are obviously reliable to exploit in comparison with inactivated and live vaccines during outbreaks to expriment in animal laboratories. This is the reason why we have been applied rTolC in the current study and the novel recombinant vaccine candidate against *S. flexneri*. The first efforts to identify *Shigella* vaccine candidate by immunoproteomics analysis was performed against *S. flexneri 2a2457T (34)*.

In the light of previous studies in *Shigella* vaccine development (17, 32, 35). Evidently, the construction of individual protein vaccines based on the outer membrane proteins against different pathogenic bacteria is efficacious.

In the current study, the TolC protein was firstly analyzed for transmembrane domains feature, homoloy and conservation properties, antigenicity, solubility, and epitope prediction by in silico. The results depicted that the TolC from *S. flexneri* ATCC12022 was highly conserved among *Shigella* species. It predicts immunization induced by TolC as the universal vaccin candidate could be effective against shigellosis causes by *S. flexneri*, *S. sonnei*, *S.flexneri*, *S. dysenteriae*.

In the vaccine studies to develop broad-spectrum vaccine candidate, homology of selected protein with mouse and human proteome is vital to determine of auto-immune response (19, 36). In the current study, TolC protein was predicted as a high conserved virulence factor in *Shigella* species and without homology with mouse and human.

Finally, TolC protein was selected and predicted through bioinformatic tools as an antigenic sequence that would induce immunogenicity in the animal laboratory.

In previous study by Baseer et al.(12), to predicts a vaccine candidate against *S. sonne,* the phicycochemical characterizations and allergenicity of TolC protein was analyzed using expacy and SORTALLER server, respectively and finally was selected as an antigenic protein. In their study, also transmembrane domains in TolC protein was absent. In present study, phisycochemical characterization of TolC protein of *S. flexneri* 12022 was predicted by Vaxigen and ANTIGENPro (http://www.ddg-pharmfac.net/vaxijen/VaxiJen/VaxiJen.html, https://scratch.proteomics.ics.uci.edu/ ) also, was without transmembrane domain and applied for next analyses.

Leow et al., through the reverse vaccinology method, found five outer membrane protein of *S. flexneri* as new vaccine candidates. In their study, outer membrane proteins just were evaluated and characterized by bioinformatic tools (37). Whereas, we used of more servers and validation steps in our study, also in comparison to their study, our in silico results were validated experimentally in animal model.

OmpA, VirG of *Shigella* were identified and evaluated experimentally as vaccine candidates before (33)(Pore et al., 2013 ;Yagnik et al., 2019, Collins et al., 2008).

The B- T cells epititope prediction is the crucial step to confirm accessibility, antigenicity, flexibility, hydrophilicity as an immunogenic protein (Ferrante, 2013).

The evaluation of the host immune response against *S. flexneri* by bioinformatics tools includes the upregulation of a series of immune cells through innate, humoral and cellular immune systems against *Shigella*.

The previous studies illustrated, the humoral immune response has been a key role in protective immunity against *Shigella*. Also, T cells were proliferated in a patient infected by *S. flexneri*. To evaluate the immunogenic potency of pre-selected proteins, B- and T –cell epitopes were evaluated by online servers (38, 39).

The scoring systems of epitope prediction demonstrate TolC has epitope density with qualified protein length and the epitope score was computed and defined and must be validated in the experimental stage. TolC is a critical member of the efflux pumps system in Gram-negative bacteria, providing broad-spectrum resistance and is also known as a multidrug-resistant agent.

In this study, the TolC protein was contained several B- T cells immunogenic epitopes. The sequence of *tolC* sequence was cloned and expressed at a size of almost 36.4 KDa by SDS-PAGE. Concentration of the purified rTolC protein was measured 25µg/µl for injection routes (mouse/group).

The immunoassay determined that the rTolC protein is immunogenic in mice. One month after the last immunization with rTolC, sera against rTolC evaluated by Insirect-ELISA assay, and the test groups showed a high level of IgG antibodies production in comparison with the PBS group. In this study, the levels of IgG titer were measured after each injection by indirect ELISA test with anti-IgG. The high levels of IgG were detectable in the sera of test mice groups after the third administration (second booster) compared to two before administrations and the control group by indirect ELISA. Therefore, TolC is capable to stimulate the humoral immune system as an antigenic exposable recombinant protein.

Also after the challenge stage, the inhibitory effect of produced antibodies against *S. flexneri* suggests TolC as a commendable antigen as a vaccine candidate.

Injection of different doses of *S. flexneri* through the intraperitoneal route led to successful induction of shigellosis infection in the mice group that received $LD_{50} = 1 \times 10^7$, depicting 50 percent of mortality. In the current study, after intraperitoneal administration of mice, results of the injection of a bacterial dose of $2 \times LD_{50}$ and $4 \times LD_{50}$ showed 90% and 75 % protection, respectively. On the other hand, after the challenge step, passive immunization resulted in extremely significant difference in survival rates between test group and control groups after 15 days post-challenge.

In conclusion, rTolC as an outer membrane protein was highly conserved among *S. flexneri*, *S. dysentriae*, *S. sonnei*, *S. boydii*. Linear epitopes and discontinuous B cell epitopes were confirmed the production of antibody in immunization of mice by rTolC. Mice immunized with rTolC antigen were potently protected against $2 \times LD_{50}$ and $4 \times LD_{50}$ with E. coli.

We believe, taking into consideration the results of in silico and in vivo approaches, rTolC applied acceptable immunogenicity as a novel vaccine against shigellosis.

The vaccine development owes to animal models historically. Primate models have the most similarity to human immune mechanisms although, the mouse models are cost-efficient and easy to handle in comparison with primate models and this model is widespread used in *Shigella* vaccine studies.

The verification of immunogenicity through the animal phase exhibiting the reverse vaccinology approach can be a cutting-edge and valid method for the vaccine development of this recombinant protein. Therefore, reverse vaccinology as the powerful and significant in silico prediction approach to identify new protein-based vaccine, reduce the costs and time of experiments and verifying vaccine candidate.

## Ethical approval

## Acknowledgements

This work was supported by the Department of Bacteriology and Virology, School of Medicine, Shiraz University of Medical Sciences, Shiraz, Iran.

## References:

1.	Sur D, Ramamurthy T, Deen J, Bhattacharya S. Shigellosis: challenges & management issues. Indian Journal of Medical Research. 2004;120(5):454.
2.	Keusch GT. Shigellosis. Bacterial infections of humans: epidemiology and control. 2009:699-724.
3.	Philpott DJ, Edgeworth JD, Sansonetti PJ. The pathogenesis of Shigella flexneri infection: lessons from in vitro and in vivo studies. Philosophical Transactions of the Royal Society of London Series B: Biological Sciences. 2000;355(1397):575-86.
4.	Puzari M, Sharma M, Chetia P. Emergence of antibiotic resistant Shigella species: A matter of concern. Journal of infection and public health. 2018;11(4):451-4.
5.	Avakh Majalan P, Hajizade A, Nazarian S, Pourmand MR, Amiri Siyavoshani K. Investigating the prevalence of Shigella species and their antibiotic resistance pattern in children with acute diarrhea referred to selected hospitals in Tehran, Iran. Journal of Applied Biotechnology Reports. 2018;5(2):70-4.
6.	Shahin K, Zhang L, Bao H, Hedayatkhah A, Soleimani-Delfan A, Komijani M, et al. An in-vitro study on a novel six-phage cocktail against multi-drug resistant-ESBL Shigella in aquatic environment. Letters in Applied Microbiology. 2021;72(3):231-7.
7.	Zamanlou S, Ahangarzadeh Rezaee M, Aghazadeh M, Ghotaslou R, Babaie F, Khalili Y. Characterization of integrons, extended-spectrum β-lactamases, AmpC cephalosporinase, quinolone resistance, and molecular typing of Shigella spp. from Iran. Infectious Diseases. 2018;50(8):616-24.
8.	Herrera CM, Schmitt JS, Chowdhry EI, Riddle MS. From Kiyoshi Shiga to Present-Day Shigella Vaccines: A Historical Narrative Review. Vaccines. 2022;10(5):645.
9.	MacLennan CA, Grow S, Ma L-f, Steele AD. The Shigella vaccines pipeline. Vaccines. 2022;10(9):1376.
10.	Sunita, Sajid A, Singh Y, Shukla P. Computational tools for modern vaccine development. Human vaccines & immunotherapeutics. 2020;16(3):723-35.
11.	Kazi A, Chuah C, Majeed ABA, Leow CH, Lim BH, Leow CY. Current progress of immunoinformatics approach harnessed for cellular-and antibody-dependent vaccine design. Pathogens and global health. 2018;112(3):123-31.
12.	Baseer S, Ahmad S, Ranaghan KE, Azam SS. Towards a peptide-based vaccine against Shigella sonnei: A subtractive reverse vaccinology based approach. Biologicals. 2017;50:87-99.

13. Kelly DF, Rappuoli R, editors. Reverse vaccinology and vaccines for serogroup B Neisseria meningitidis. Hot Topics in Infection and Immunity in Children II; 2005: Springer.
14. Du D, Wang Z, James NR, Voss JE, Klimont E, Ohene-Agyei T, et al. Structure of the AcrAB–TolC multidrug efflux pump. Nature. 2014;509(7501):512-5.
15. Heinson AI, Woelk CH, Newell M-L. The promise of reverse vaccinology. International health. 2015;7(2):85-9.
16. Gardy JL, Laird MR, Chen F, Rey S, Walsh C, Ester M, et al. PSORTb v. 2.0: expanded prediction of bacterial protein subcellular localization and insights gained from comparative proteome analysis. Bioinformatics. 2005;21(5):617-23.
17. Hajialibeigi A, Amani J, Gargari SLM. Identification and evaluation of novel vaccine candidates against Shigella flexneri through reverse vaccinology approach. Applied Microbiology and Biotechnology. 2021;105:1159-73.
18. Talukdar S, Zutshi S, Prashanth K, Saikia KK, Kumar P. Identification of potential vaccine candidates against Streptococcus pneumoniae by reverse vaccinology approach. Applied biochemistry and biotechnology. 2014;172:3026-41.
19. Rashid MI, Naz A, Ali A, Andleeb S. Prediction of vaccine candidates against Pseudomonas aeruginosa: An integrated genomics and proteomics approach. Genomics. 2017;109(3-4):274-83.
20. Doytchinova IA, Flower DR. Identifying candidate subunit vaccines using an alignment-independent method based on principal amino acid properties. Vaccine. 2007;25(5):856-66.
21. Hattotuwagama CK, Guan P, Doytchinova IA, Zygouri C, Flower DR. Quantitative online prediction of peptide binding to the major histocompatibility complex. Journal of Molecular Graphics and Modelling. 2004;22(3):195-207.
22. León Y, Zapata L, Molina RE, Okanovič G, Gómez LA, Daza-Castro C, et al. Intranasal immunization of mice with multiepitope chimeric vaccine candidate based on conserved autotransporters SigA, Pic and Sap, confers protection against Shigella flexneri. Vaccines. 2020;8(4):563.
23. Sharma M, Dixit A. Identification and immunogenic potential of B cell epitopes of outer membrane protein OmpF of Aeromonas hydrophila in translational fusion with a carrier protein. Applied microbiology and biotechnology. 2015;99:6277-91.
24. Chen R. Bacterial expression systems for recombinant protein production: E. coli and beyond. Biotechnology advances. 2012;30(5):1102-7.
25. Frey A, Di Canzio J, Zurakowski D. A statistically defined endpoint titer determination method for immunoassays. Journal of immunological methods. 1998;221(1-2):35-41.
26. Saha S, Raghava GPS. Prediction of continuous B-cell epitopes in an antigen using recurrent neural network. Proteins: Structure, Function, and Bioinformatics. 2006;65(1):40-8.
27. Guan P, Doytchinova IA, Zygouri C, Flower DR. MHCPred: bringing a quantitative dimension to the online prediction of MHC binding. Applied bioinformatics. 2003;2(1):63-6.
28. Schnupf P, Sansonetti PJ. Shigella pathogenesis: new insights through advanced methodologies. Bacteria and Intracellularity. 2019:15-39.
29. Esmailnia E, Amani J, Gargari SLM. Identification of novel vaccine candidate against Salmonella enterica serovar Typhi by reverse vaccinology method and evaluation of its immunization. Genomics. 2020;112(5):3374-81.
30. Rauta PR, Ashe S, Nayak D, Nayak B. In silico identification of outer membrane protein (Omp) and subunit vaccine design against pathogenic Vibrio cholerae. Computational biology and chemistry. 2016;65:61-8.
31. Chitradevi S, Kaur G, Sivaramakrishna U, Singh D, Bansal A. Development of recombinant vaccine candidate molecule against Shigella infection. Vaccine. 2016;34(44):5376-83.

32. Sharma D, Yagnik B, Baksi R, Desai N, Padh H, Desai P. Shigellosis murine model established by intraperitoneal and intranasal route of administration: a comparative comprehension overview. Microbes and Infection. 2017;19(1):47-54.
33. Pore D, Chakrabarti MK. Outer membrane protein A (OmpA) from Shigella flexneri 2a: a promising subunit vaccine candidate. Vaccine. 2013;31(36):3644-50.
34. Jennison AV, Raqib R, Verma NK. Immunoproteome analysis of soluble and membrane proteins of Shigella flexneri 2457T. World Journal of Gastroenterology: WJG. 2006;12(41):6683.
35. Yagnik B, Sharma D, Padh H, Desai P. Oral immunization with LacVax® OmpA induces protective immune response against Shigella flexneri 2a ATCC 12022 in a murine model. Vaccine. 2019;37(23):3097-105.
36. Vishnu US, Sankarasubramanian J, Gunasekaran P, Rajendhran J. Identification of potential antigens from non-classically secreted proteins and designing novel multitope peptide vaccine candidate against Brucella melitensis through reverse vaccinology and immunoinformatics approach. Infection, Genetics and Evolution. 2017;55:151-8.
37. Leow CY, Kazi A, Ismail CMKH, Chuah C, Lim BH, Leow CH, et al. Reverse vaccinology approach for the identification and characterization of outer membrane proteins of Shigella flexneri as potential cellular-and antibody-dependent vaccine candidates. Clinical and experimental vaccine research. 2020;9(1):15-25.
38. Ashida H, Mimuro H, Sasakawa C. Shigella manipulates host immune responses by delivering effector proteins with specific roles. Frontiers in Immunology. 2015;6:219.
39. Mani S, Toapanta FR, McArthur MA, Qadri F, Svennerholm A-M, Devriendt B, et al. Role of antigen specific T and B cells in systemic and mucosal immune responses in ETEC and Shigella infections, and their potential to serve as correlates of protection in vaccine development. Vaccine. 2019;37(34):4787-93.

**Additional Supplementaries**

**Graphical Abstract**

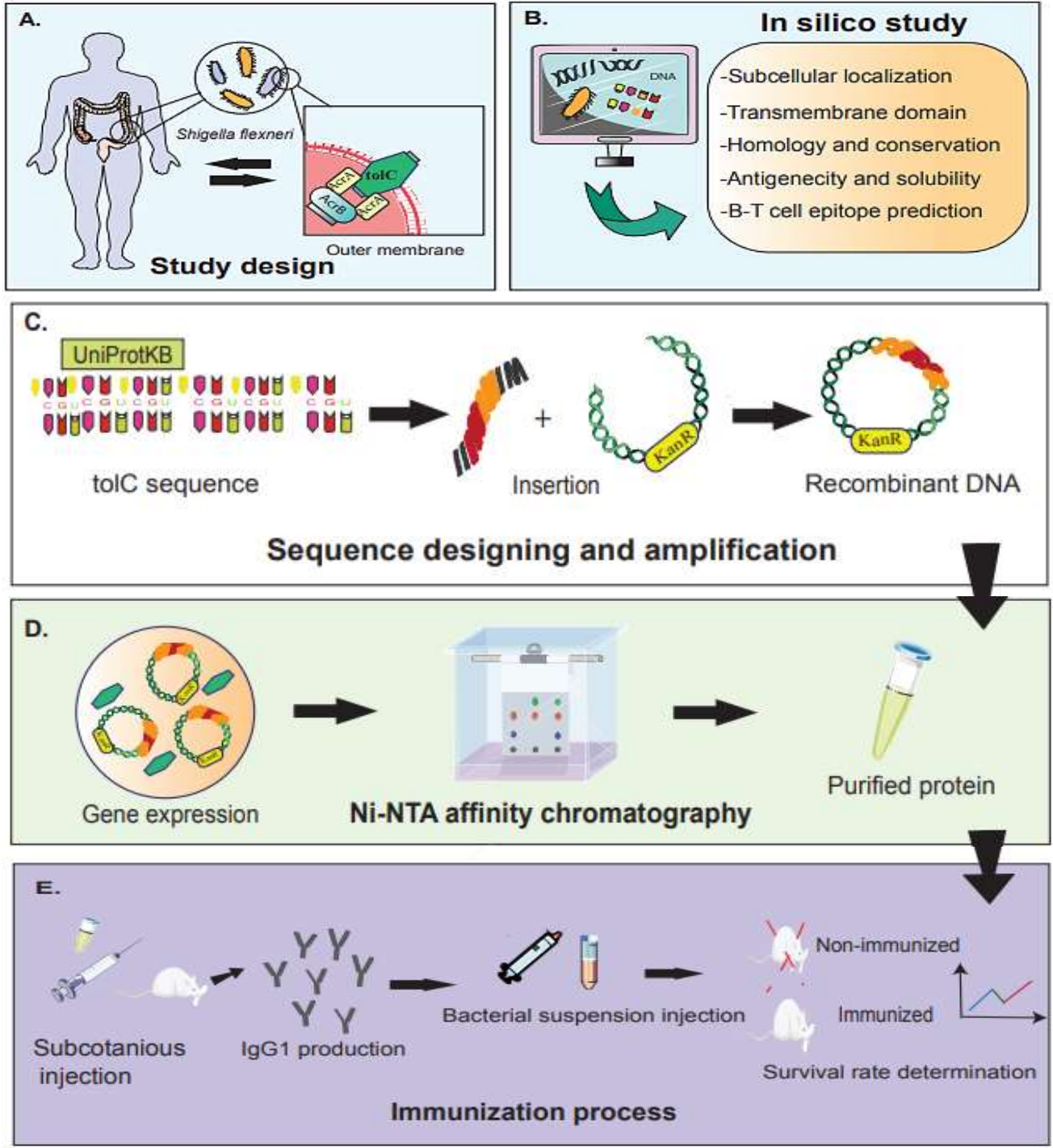

Parts of the figure were drawn by using pictures from Servier Medical Art. Sevier Medical Art by sevier is licensed under a Creative Commons Attribution 3.0 Unported License. https://smart.servier.com/

**tolC Gene sequence:**

```
Created: 2021Äê9ÔÂ14ÈÕ ÐÇÆÚ¶þ 9:28
^^
gaattcCGCAAAAGCGCAGCAGATCGCGATGCAGCATTTGAAAAAATTAATGAAGCACGTAGCCCGCTGC
TGCCGCAGCTGGGTCTGGGTGCAGATTATACCTATAGCAATGGTTATCGTGATGCAAATGGTATTAATAG
CAATGCAACCAGCGCAAGCCTGCAGCTGACCCAGAGCATTTTTGATATGAGCAAATGGCGTGCACTGACC
CTGCAGGAAAAAGCAGCAGGTATTCAGGATGTTACCTATCAGACCGATCAGCAGACCCTGATTCTGAATA
CCGCAACCGCATATTTTAATGTTCTGAATGCAATTGATGTTCTGAGCTATACCCAGGCACAGAAAGAAGC
AATTTATCGTCAGCTGGATCAGACCACCCAGCGTTTTAATGTTGGTCTGGTTGCAATTACCGATGTTCAG
AATGCACGTGCACAGTATGATACCGTTCTGGCAAATGAAGTTACCGCACGTAATAATCTGGATAATGCAG
TTGAACAGCTGCGTCAGATTACCGGTAATTATTATCCGGAACTGGCAGCACTGAATGTTGAAAATTTTAA
AACCGATAAACCGCAGCCGGTTAATGCACTGCTGAAAGAAGCAGAAAAACGTAATCTGAGCCTGCTGCAG
GCACGTCTGAGCCAGGATCTGGAACGTGAACAGATTCGTCAGGCACAGGATGGTCATCTGCCGACCCTGG
ATCTGACCGCAAGCACCGGTATTAGCGATACCAGCTATAGCGGTAGCAAAACCCGTGGTGCAGCAGGTAC
CCAGTATGATGATAGCAATATGGGTCAGAATAAAGTTGGTCTGAGCTTTAGCCTGCCGATTTATCAGGGT
GGTATGGTTAATAGCCAGGTTAAACAGGCACAGTATAATTTTGTTGGTGCAAGCTAAaagctt
```